\documentclass{article}
\usepackage[preprint]{spconfa4}
\usepackage{amsmath,graphicx}
\usepackage{comment}
\usepackage{eucal}
\usepackage{multirow}
\usepackage{cite}

\copyrightnotice{978-1-6654-6867-1/22/\$31.00~\copyright2022 IEEE}
                   
\title{Realistic sources, receivers and walls improve the generalisability of virtually-supervised blind acoustic parameter estimators}
\name{Prerak Srivastava, Antoine Deleforge, Emmanuel Vincent}
\address{Universit\'e de Lorraine, CNRS, Inria, Loria, F-54000 Nancy, France\\ \{prerak.srivastava, antoine.deleforge, emmanuel.vincent\}@inria.fr\\  }
\begin{document}
%
\maketitle
\begin{abstract}

Blind acoustic parameter estimation consists in inferring the acoustic properties of an environment from recordings of unknown sound sources. Recent works in this area have utilized deep neural networks trained either partially or exclusively on simulated data, due to the limited availability of real annotated measurements. In this paper, we study whether a model purely trained using a fast image-source room impulse response simulator can generalize to real data. We present an ablation study on carefully crafted simulated training sets that account for different levels of realism in source, receiver and wall responses. The extent of realism is controlled by the sampling of wall absorption coefficients and by applying measured directivity patterns to microphones and sources. A state-of-the-art model trained on these datasets is evaluated on the task of jointly estimating the room's volume, total surface area, and octave-band reverberation times from multiple, multichannel speech recordings. Results reveal that every added layer of simulation realism at train time significantly improves the estimation of all quantities on real signals.




\end{abstract}
\section{Introduction}
\label{sec:intro}
Learning-based approaches have become ubiquitous in nearly all areas of audio signal processing. This includes the task of blind acoustic parameter estimation \cite{microsoft_rt60,genovese2019blind,bryan2020impulse,srivastava2021blind,gotz2022blind}, which has applications in the automatic adaptation of augmented reality devices \cite{valimaki2015assisted} or hearing aids \cite{kates2001room} to their environment. A general bottleneck of learning-based methods is the need for large annotated training datasets. A widely used technique to emulate sound scenes in a variety of acoustic environments is to convolve dry source signals with room impulse responses (RIRs). A RIR is defined as the linear time-domain response of a microphone to an impulse sound source inside a room. It is determined by the combined geometrical and acoustical properties of the source-microphone-room system.

Because acquiring and annotating the tens of thousands of RIRs needed to train typical deep neural network models is impractical, a number of data augmentation techniques have been proposed in recent years. In the context of automatic speech recognition, the authors of \cite{szoke2019building} showed that using a small set of carefully selected measured RIRs that match the target environment conditions can yield results comparable to using a large set of simulated RIRs. In  \cite{microsoft_rt60,genovese2019blind,gotz2022blind}, it was shown that combining real and synthetic RIRs improves the generalization of blind room volume and blind reverberation time estimators. In \cite{bryan2020impulse}, a data augmentation scheme was proposed to increase the size and diversity of a real RIR dataset.
While these studies show that leveraging a few hundred annotated real RIRs for training can be effective, such real data are not always available depending on the task at hand, and are costly to acquire even in relatively small amount.

An alternative is to train models on synthetic data only, an approach referred to as \textit{virtually supervised learning} \cite{gaultier2017vast}. The authors of \cite{gaultier2017vast} released a training set of $\sim$100k RIRs generated with a hybrid simulator combining the image-source method (ISM) \cite{allen1979image} and ray tracing \cite{schimmel2009fast} for modeling diffusion, and used it to train a binaural 3D sound source localization model which generalized better to real data than a model trained on anechoic head-related transfer functions. On a larger scale, the authors of \cite{tang2022gwa} recently released a dataset of $\sim$2 million high quality synthetic RIRs. They leveraged realistic 3D house models, an automatically matched database of material absorption profiles, and a highly accurate acoustic simulator combining a finite-difference time-domain wave equation solver at low frequencies with ray tracing at high frequencies. The models trained on this dataset outperformed models trained with less realistic simulators on a variety of single-channel speech processing tasks. Generating such large RIR datasets is much more computationally demanding than using the classical ISM for shoe-box rooms, as done in \cite{genovese2019blind,bryan2020impulse,gotz2022blind,gaultier2017vast}, namely, $\sim$1300 CPU/GPU hours for \cite{tang2022gwa} vs.\ $\sim$700 CPU hours for \cite{gaultier2017vast}. Despite this, these datasets do not straightforwardly generalize to, e.g., multichannel settings with a specific microphone array geometry. To alleviate such stringent computational requirements, a fast stochastic RIR simulator was proposed in \cite{masztalski2020storir} and used to train speech enhancement models. While better generalization to real data was demonstrated w.r.t.\ the ISM, the approach cannot be used in multichannel settings nor for geometric parameter estimation.

In this paper, we study how well a model purely trained on various extensions of the ISM generalizes to real data. Specifically, we perform an ablation study on 7 simulated training sets to assess the impact of using different source and/or receiver directivity profiles and wall absorption coefficient distributions. These extensions will be incorporated in a future release of the open source Pyroomacoustics simulator \cite{scheibler2018pyroomacoustics}. We focus here on the task of jointly estimating the room's volume, total surface area and reverberation time per octave band using multiple, multichannel speech recordings, using our recently proposed state-of-the art model \cite{srivastava2021blind}. The model is tested on real reverberant speech signals from the dEchorate dataset \cite{carlo2021dechorate} corresponding to 6 distinct microphone arrays, 7 distinct loudspeakers and 4 distinct room configurations. The results reveal that every added layer of realism at train time significantly and consistently improves the estimation of all quantities at test time, while keeping the computational cost of simulations similar to the classical ISM.

\section{SIMULATION}
\label{sec:simulation}
Reverberated speech can be simulated  in the discrete-time domain as follows :
\begin{equation}
\label{eq:signal_model}
x[t] = (h * s)[t] + n[t]    
\end{equation}
where $*$ denotes discrete convolution, $h$ the RIR from the source to the microphone, $s$ the dry source signal, and $n$ additive noise. One of the most widely used method to simulate $h$ is the ISM \cite{allen1979image} due to its simplicity and unmatched computational efficiency, especially in shoe-box rooms. It models $h$ as the sum of $K$ free field responses to \textit{image sources} (IS) whose positions correspond to iterated spatial reflections of the original source with respect to the room's boundaries. While the original method
only considered rigid boundaries and omnidirectional sources and receivers, it can be extended to account for directivities and frequency-dependent responses with little computational overhead. A general formula in the Fourier domain can be written as follows (see, e.g., \cite[Chap. 5]{schroder2011physically}):
\begin{align}
    \hat{h}(f)=\sum_{k=1}^{K}\,\frac{e^{-j 2 \pi f \tau_{k}/F_{\textrm{s}}}}{c\tau_{k}} \,d_{\textrm{air}}(f)\,d_{k}(f)\,&\hat{g}_{\textrm{src}}(\theta_{k}^{\textrm{out}},\phi_{k}^{\textrm{out}},f) \nonumber\\
    &\hat{g}_{\textrm{mic}}(\theta_{k}^{\textrm{in}},\phi_{k}^{\textrm{in}},f)
    \label{eq:transfer_function}
\end{align}
where $\hat{h}(f)$ is a frequency-domain representation of $h$ with $f$ in Hz, $c$ denotes the speed of sound in $\textrm{m}/\textrm{s}$, $F_\textrm{s}$ is the frequency of sampling in Hz, $\tau_k$ is the time of arrival of IS $k$ at the microphone in s, $d_{\textrm{air}}(f)$ denotes atmospheric attenuation, 
$d_{k}(f)$ is the damping coefficient of IS $k$ due to surface absorption, $\hat{g}_{\textrm{src}}$ and $\hat{g}_{\textrm{mic}}$ respectively denote the source and microphone directivity reponses and $(\theta_{k}^{\textrm{out}},\phi_{k}^{\textrm{out}})$ and $(\theta_{k}^{\textrm{in}},\phi_{k}^{\textrm{in}})$ respectively denote the azimuth and elevation angles of departure and arrival of image source $k$ towards the microphone. We have $\tau_k=|| \boldsymbol{r}_{\textrm{mic}} - \boldsymbol{r}_{k} ||/c$ where $\boldsymbol{r}_{\textrm{mic}}$ and $\boldsymbol{r}_{k}$ denote the positions of the microphone and of IS $k$, respectively. In practice, each term in the sum (\ref{eq:transfer_function}) can be calculated over $F$ discrete positive frequencies using the fractional part of $\tau_k$ only, before adding their $2F$-point inverse discrete Fourier transform to $h$ with a delay equal to the integral part of $\tau_k$. Despite the availability of many open-source ISM simulators, none of them implement (\ref{eq:transfer_function}) in its full generality, to the best of our knowledge. We plan to publish our implementation as part of a future release of the Pyroomacoustics simulator \cite{scheibler2018pyroomacoustics}.

\label{sec:format}
\subsection{Source and Receiver Directivities}

Source and receiver directivities have a strong impact on RIRs, as highlighted in, e.g., \cite{knuttel2013influence}
. Despite this, the vast majority of simulation-based supervised models in the audio literature neglect this effect by considering ideal omnidirectional sources and receivers, namely, 
\begin{equation} \tag{$\mathcal{O}$}
\label{eq:omni}
  \hat{g}_{\textrm{src}}(\theta,\phi,f) = \hat{g}_{\textrm{mic}}(\theta,\phi,f)=1.
\end{equation}
A first way to account for directivities is to use the following analytical formula with parameter $\beta\in[0,1]$:
\begin{equation} \tag{$\mathcal{A}_\beta$}
    \label{eq:analytical_formula}
    \hat{g}(\theta,\phi,f)= \hat{g}(\theta) = \beta + (1-\beta)\cos(\theta).
\end{equation}
Setting $\beta$ to $0, 0.25, 0.5, 0.75$ and $1$ respectively yields figure-of-eight, hypercardioid, cardioid, subcardioid and omnidirectional patterns. Note that this formula neglects elevation dependency and, more crucially, frequency dependency. Yet, the directivity patterns of real sources and receivers tend to be nearly omnidirectional in lower frequencies and directive and concentrated in higher frequencies.

To account for this, we propose to use measured ($\mathcal{M}$) directivity patterns from the DIRPAT dataset \cite{brandner2018dirpat}. This dataset includes measurements for a variety of loudspeakers and guitar amplifiers, a Br\"{u}el \& Kj\ae r head and torso mouth simulator, an AKG C414 microphone in four settings (Omni, Cardiod, Supercardiod, F-8) and 3 other microphones. For each source and receiver, the measurements are available as time-domain finite impulse responses on a discrete spherical grid $G$ with 30 regularly-spaced azimuth and 16 or 18 regularly-spaced elevations, $\{g(\tilde{\theta}_{j},\tilde{\phi}_{j},t)\}_{j\in G}$.
To obtain a better covering of incidence and exit angles in simulations, we use spherical harmonic interpolation to map the DFT of each impulse response at each $f$ to a denser and more uniform Fibonacci grid $G'$ with $1,000$ points, yielding
$\{\hat{g}(\tilde{\theta}'_{j},\tilde{\phi}'_{j},f)\}_{j\in G'}$. Interpolation is implemented using Voronoi-cell-based weighted least squares as described in \cite[Chap. 4]{zotter2009analysis}. The final gain for IS $k$ with continuous exit and incidence angles is obtained by nearest neighbour interpolation over $G'$.

\subsection{Absorption Profiles}
Calculating the damping coefficients $d_k(f)$ in (\ref{eq:transfer_function}) requires us to define the absorption profiles of the 6 surfaces inside the shoe-box room. Since the absorption coefficients of materials are usually provided in 6 octave bands $b\in B=\{125, 250, 500, 1\textrm{k}, 2\textrm{k}, 4\textrm{k}$\} Hz, we use the following interpolation formula:
\begin{equation}
    \label{eq:octave_band}
    d_{k}(f)=\textstyle\sum_{b}\tilde{d}_{k}(b)\nu_{b}(f)
\end{equation}
where $\nu_{b}(f)$ denotes the positive half-cosine octave band filters used in the Pyroomacoustics toolbox \cite{scheibler2018pyroomacoustics}. Note that the same formula is used to obtain $d_{\textrm{air}}(f)$ from measured atmospheric attenuation coefficients. We further have
\begin{equation}
    \tilde{d}_{k}(b)=\textstyle\prod_{j}\rho_{i_{k,j}}(b) \quad \textrm{and} \quad
    \rho_{i}(b)=\sqrt{1-\alpha_{i}(b)},
\end{equation}
where $\alpha_{i}(b)$ is the absorption coefficient of surface $i$, $\rho_{i}(b)$ is its reflection coefficient, and $i_{k,j}$ is the index of the $j$-th surface encountered by IS $k$.
We consider two different sampling strategies for absorption coefficients. The naive one ($\mathcal{N}$) uses a single frequency-independent coefficient $\alpha_i(b)=\alpha\in[0.02,0.5]$ for all surfaces\footnote{This range was chosen to make $\textrm{RT}_{60}$ distributions similar for $\mathcal{N}$ and $\mathcal{R}\mathcal{B}$.}, drawn uniformly at random for each simulated room. The second, called \textit{reflectivity-biased} ($\mathcal{R}\mathcal{B}$) is inspired by the one proposed in \cite{foy2021mean}. First, a number of \textit{reflective} surfaces from 0 to 6 is picked uniformly at random. Each reflective surface is then given a frequency-independent absorption coefficient drawn uniformly at random in $[0.01,0.12]$. Then, each remaining surface is randomly assigned to be either a wall, a ceiling or a floor. Its absorption coefficients per octave bands are then drawn uniformly at random inside ranges defined according to material databases of the assigned type, as detailed in \cite{foy2021mean}.
Finally, note that the interpolation formula (\ref{eq:octave_band}) currently used in Pyroomacoustics yields surface responses that are \textit{zero phase}, \textit{i.e.}, non-causal and with potentially large delays. To correct this in $\mathcal{R}\mathcal{B}$, we converted the positive responses given by (\ref{eq:octave_band}) to \textit{minimum phase} filters, that are by definition causal and with fastest possible decay.
\begin{figure}[t]
  \centering
  \centerline{\includegraphics[width=8.5cm]{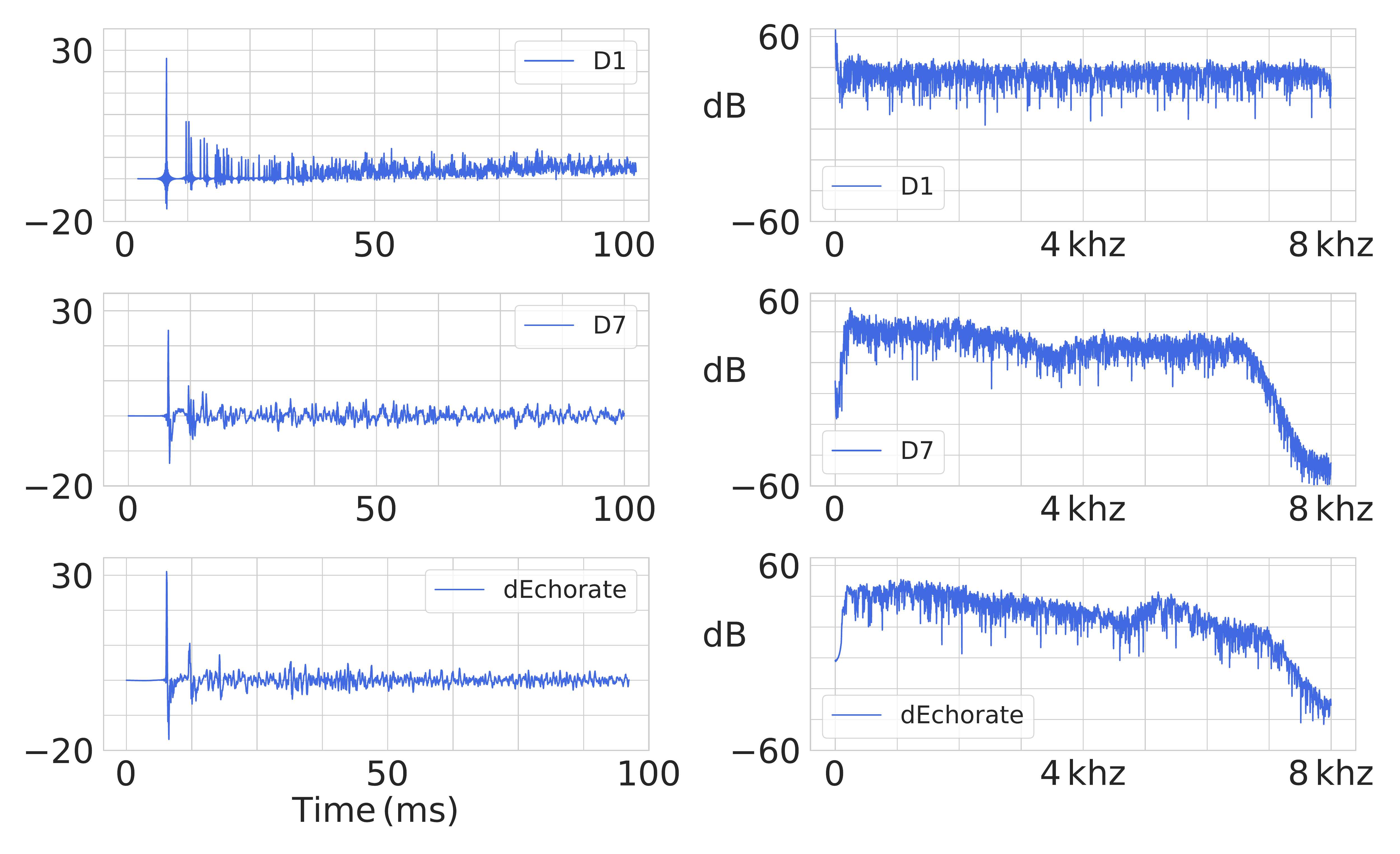}}
  \caption{Normalized room impulse responses (left) and their magnitude frequency responses (right) from simulated datasets D1 (top), D7 (middle) and the real dataset dEchorate \cite{carlo2021dechorate} (bottom).}
  \label{fig:rirs}
\end{figure}

\section{Task, Model and Training Sets}
\vspace{-3mm}
The task evaluated in this paper is similar to the one presented in \cite{srivastava2021blind}. Given a set of $P$ two-channel noisy speech recordings made at different source-receiver positions in a room, the goal is to jointly estimate the room's volume $V$, total surface area $S$, and octave-band-wise reverberation times $\{\textrm{RT}_{60}(b)\}_{b\in B}$. Contrary to \cite{srivastava2021blind}, we do not estimate mean absorption coefficients here due to the lack of annotated real data for evaluation. Throughout this study, we use $P=3$ speech recordings of 3 seconds each made with a 2-element linear array of aperture $22.5$~cm placed parallel to the floor at 3 different positions and a fixed source position per room.
To perform the task, we use the neural network model in \cite{srivastava2021blind}, which combines single-channel and inter-channel input features through several convolution, pooling and dense layers, and jointly estimates the means and variances of target quantities using a Gaussian maximum-likelihood loss. The input features are calculated using the short-time Fourier transform with a window size of $48$~ms and $50\%$ overlap. For training, we used the ADAM optimizer with a learning rate of $10^{-4}$, an $l_2$ regularization of $10^{-5}$ and a batch size of 16. Moreover, a dropout rate of $0.5$ as well as an $l_1$ regularization on the network's weights with $\lambda=10^{-3}$ were needed to avoid over-fitting. 

\begin{table*}[t!]
\vspace*{-1cm}
    \centering
    \footnotesize
    \begin{tabular}{c|c|c|c|c|c|c|c|c|c|c|c|c|c|c|c}
\hline
Training & \multirow{2}{*}{walls} & \multirow{2}{*}{src} & \multirow{2}{*}{mic} & \multicolumn{2}{|c|}{$\textrm{RT}_{60}$ (500 Hz)} & \multicolumn{2}{c|}{$\textrm{RT}_{60}$ (1~kHz)} & \multicolumn{2}{c|}{$\textrm{RT}_{60}$ (2~kHz)} & \multicolumn{2}{c|}{$\textrm{RT}_{60}$ (4~kHz)} & \multicolumn{2}{c|}{$S$} & \multicolumn{2}{c}{$V$} \\
\cline{5-16}
 dataset & & & & realist. & real & realist. & real& realist. & real & realist.  & real& realist. & real & realist. & real \\
\hline
D1 & $\mathcal{N}$ & $\mathcal{O}$ & $\mathcal{O}$ & $0.182$ & $0.193$ & $0.150$ & $0.160$ & $0.150$ & $0.108$ & $0.139$ & $0.185$ & $97.04$ & $71.00$ & $98.47$ & $75.68$ \\
D2 & $\mathcal{RB}$ & $\mathcal{O}$ & $\mathcal{O}$ & $0.186$ & $0.182$ & $0.189$ & $0.140$ & $0.228$ & $0.128$ & $0.226$ & $0.198$ & $69.28$ & $45.11$ & $75.56$ & $55.16$ \\
D3 & $\mathcal{RB}$ & $\mathcal{O}$ & $\mathcal{M}$ & $0.198$ & $0.115$ & $0.158$ & $\textbf{0.098}$ & $0.133$ & $0.078$ & $0.110$ & $0.156$ & $103.57$ & $52.76$ & $108.58$ & $61.82$ \\
D4 & $\mathcal{RB}$ & $\mathcal{A_{\beta}}$ & $\mathcal{O}$& $0.170$ & $0.167$ & $0.138$ & $0.134$ & $0.151$ & $0.121$ & $0.155$ & $0.197$ & $77.61$ & $37.91$ & $84.89$ & $48.95$ \\  
D5 & $\mathcal{RB}$ & $\mathcal{M}$ & $\mathcal{O}$ & $0.168$ & $0.133$ & $0.143$ & $0.112$ & $0.153$ & $\textbf{0.066}$ & $0.119$ & $0.155$ & $50.46$ & $\textbf{21.46}$ & $53.09$ & $\textbf{18.57}$ \\
D6 & $\mathcal{N}$ & $\mathcal{M}$ & $\mathcal{M}$ & $0.152$ & $0.151$  & $0.177$ & $0.133$ & $0.177$ & $0.084$ & $\textbf{0.098}$ & $0.159$ & $37.88$ & $35.88$ & $41.08$ & $31.11$ \\ 
D7 & $\mathcal{RB}$ & $\mathcal{M}$ & $\mathcal{M}$& $\textbf{0.134}$ & $\textbf{0.080}$ & $\textbf{0.105}$ & $\textbf{0.103}$ & $\textbf{0.116}$ & $\textbf{0.064}$ & $\textbf{0.092}$ & $\textbf{0.140}$ & $\textbf{25.22}$ & $32.69$ & $\textbf{28.63}$ & $30.57$ \\

\hline
\end{tabular}
    \caption{Mean absolute errors in reverberation time ($\textrm{RT}_{60}$, in s), surface ($S$, in $\textrm{m}^2$) and volume ($V$, in $\textrm{m}^3$) estimation achieved over a realistic simulated test set and a real test set using the same model trained on 7 simulated training datasets. Bold numbers indicate the best statistically significant result per column, based on 98\% confidence intervals.}
    \label{tab:real_results}
\end{table*}

The model is trained on 7 different datasets numbered D1 to D7 and incorporating the different levels of simulation realism described in Section \ref{sec:simulation}, as summarized in the first four columns of Table~\ref{tab:real_results}. A qualitative comparison between RIRs from D1, D7 and the real dataset dEchorate \cite{carlo2021dechorate} is shown in Fig.~\ref{fig:rirs}. As can be seen, RIRs from the most realistic dataset D7 match real RIRs better. For each source in D5, D6 and D7, a random directivity pattern among the Genelec 8020, Neumann KH120A and Yamaha DXR8 loudspeakers of DIRPAT is randomly oriented with pointing direction parallel to the floor. In D4, sources are oriented similarly but with the analytical directivity pattern (\ref{eq:analytical_formula}) and a random value of $\beta$ in $\{0.25,0.5,0.75\}$. For each individual receiver in D3, D6 and D7, the directivity pattern of the omnidirectional AKG C414 microphone of DIRPAT is rotated uniformly at random over the sphere. 
For each dataset, 3 two-channel RIRs are simulated at a rate of 16~kHz in 30,000 rooms with length, width and height drawn uniformly at random in $[3, 10]\times[3, 10]\times[2, 4.5]$, in meters.
The two-channel RIRs are convolved with speech excerpts from the Librispeech corpus \cite{panayotov2015librispeech} according to (\ref{eq:signal_model}). As in \cite{srivastava2021blind}, uncorrelated white Gaussian noise and diffuse speech-shaped noise convolved with the late part of a random two-channel RIR in the room are added to the reverberated signal. For each dataset, a reference speech signal corresponding to an emitter placed 1 meter away and facing a receiver in the middle of a $5 \times 5 \times3 $~m room with $\alpha=0.2$ is used to calibrate noise levels. This yields signal-to-noise ratios in the range $[15,75]$~dB across the datasets. Matching validation sets of 3,000 rooms are constructed for each training set. The model converges in 80 to 120 epochs using a patience of 15 epochs on the validation sets.

\section{Test sets and Results}

We evaluated the models trained on D1, \dots, D7 on two test sets in terms of the mean absolute error on each of the considered quantities. We excluded estimated reverberation times at 125 and 250~Hz because, consistently with \cite{srivastava2021blind,foy2021mean}, all models performed poorly on these quantities due to the fact that geometry-based simulation is inaccurate below the Schroeder frequency ($\sim$500 Hz). The first test set (\textit{realist.}) consists of 400 rooms simulated similarly to D7, but with out-of-training source directivities picked randomly among the Tannoy System 1200 and Lambda Labs CX-1A loudspeakers or the Br\"{u}el \& Kj\ae r 4128C HATS mouth simulator. The second test set (\textit{real}) contains real reverberant speech signals from the dEchorate dataset \cite{carlo2021dechorate}. These signals were recorded in a rectangular room of size $5.7\times 6 \times 2.4$~m $(V=82$~$\textrm{m}^3$, $S=125$~$\textrm{m}^2)$, whose walls, floor and ceiling were switched between an absorbent and a reflective mode. We excluded from the dataset the 6 \textit{semi-anechoic} room configurations (at most one reflective surface) due to reverberation times below $250$~ms in all bands, and the furnished room due to inconsistent results. This leaves 4 room configurations with 2 to 5 reflective surfaces and octave-band-wise reverberation times ranging from $250$ to $810$~ms. In each room, 6 two-element linear sub-arrays with aperture $22.5$~cm and 7 loudspeakers with distinct positions and orientations are available. The microphones are omnidirectional AKG CK32, 6 of the sources are directional Avanton MixCubes and the last source is a lightweight Br\"{u}el \& Kj\ae r omnidirectional loudspeaker. None of these receivers and sources are present in the directivity dataset used to build the training sets. Reverberant speech recordings are available for every source-receiver pair in each room.
For evaluation purposes, we consider all 140 possible combinations of three two-element arrays recording a fixed source in each of the 4 rooms, resulting in 560 test cases in total.

The results are shown in Table~\ref{tab:real_results}. We first see that the most realistic training set D7 yields best or second best results for all quantities on both test sets. Interestingly, the dataset D5, which is identical but with a simpler omnidirectional microphone model, results in better source and volume estimation than D7 on the real test set, suggesting that mismatched microphone responses at train and test times can degrade the accuracy. A solution could be to use more diverse microphone responses at train time. More generally, by comparing D2 to D3 on the one hand and D5 to D6 on the other hand, we see that using measured microphone directivities significantly improves $\textrm{RT}_{60}$ estimation over both test sets, but tends to degrade $S$ and $V$ estimation.
By comparing the esults obtained using D1 and D2 on the one hand and D6 and D7 on the other hand, we see that the more realistic \textit{reflectivity-biased} ($\mathcal{R}\mathcal{B}$) sampling strategy for wall absorption coefficients significantly and consistently outperforms the naive scheme ($\mathcal{N}$) across all target quantities and both test sets. Finally, by comparing results obtained using D2, D4 and D5, we observe that increasing the realism of source directivity consistently improve results across all target quantities and both test sets. This trend is dramatically confirmed by comparing the results obtained with D3 and D7, especially in geometry estimation.

\section{CONCLUSION}
\label{sec:majhead}
We studied the impact of carefully crafted simulated training sets including different layers of realism in source, receiver and wall responses on the generalization of a blind acoustic parameter estimator. Although the literature on learning-based audio signal processing tends to neglect these aspects, our results demonstrate that they can significantly and consistently improve generalization to real data, with a limited computational overhead. In particular, employing measured source directivities and reflectivity-biased wall sampling reduced errors on room geometry estimation by over $70\%$ compared to the same model trained on a naive dataset, over a variety of real acoustic scenes.
Future work may include a finer study on the impact of microphone responses as well as the impact of noise and acoustic diffusion. Generalization of this work to larger real datasets and other multichannel tasks, e.g., sound source localization, will also be explored.

\section{Acknowledgments}
This work was made with the support of ANR through project HAIKUS ``Artifical Intelligence applied to augmented acoustic scenes'' (ANR-19-CE23-0023). Experiments presented in this paper were carried out using the Grid’5000 testbed, supported by a scientific interest group hosted by Inria and including CNRS, RENATER and several Universities as well as other organizations (see https://www.grid5000.fr).

\bibliographystyle{IEEEbib}
\bibliography{refs}

\end{document}